\documentclass[onecolum,secnumarabic,amssymb,nobibnotes,longbibliography,preprint]{revtex4-1}

\usepackage{graphicx,color,amsmath}
\usepackage[normalem]{ulem}

\begin{document}

\title{Casimir scaling in glueballs in SU($N$) and Sp($2N$) gauge theories: hints from constituent approaches}

\author{Fabien Buisseret$^{1,2}$}
\email[e-mail:]{fabien.buisseret@umons.ac.be}
\author{Cyrille Chevalier$^1$}
\email[e-mail:]{cyrille.chevalier@umons.ac.be}
\author{Vincent Mathieu$^3$}
\email[e-mail:]{vmathieu@ub.edu}
\author{Claude Semay$^1$}
\email[e-mail:]{claude.semay@umons.ac.be}
	\affiliation{$^1$ Service de Physique Nucléaire et Subnucléaire, UMONS Research Institute for Complex Systems, Université de Mons, 20 Place du Parc, 7000 Mons, Belgium}
	\address{$^2$ CeREF, 159 Chauss\'ee de Binche, 7000 Mons, Belgium}
	\address{$^3$ Departament de F\`isica Qu\`antica i Astrof\`isica and Institut de C\`iencies del Cosmos, Universitat de Barcelona, E-08028 Barcelona, Spain}

\begin{abstract}
		We show that the lattice glueball masses $M_G$ versus $N$ in SU($N$) and Sp($2N$) Yang-Mills theories scale as $\frac{M_G}{\sqrt\sigma}\sim \sqrt{\frac{C_2(adj)}{C_2(f)}}$, with $\sigma$ the fundamental string tension and $C_2(adj)$ and $C_2(f)$ the quadratic Casimir of the gauge algebra in the adjoint and fundamental representations. This scaling behaviour is followed by the great majority of available lattice glueball states, and may set constraints on $SU(3)$ models by imposing a specific behaviour at $N\neq 3$. The observed scaling is compatible with two assumptions: (1) The glueball masses are proportional to the square root of the adjoint string tension, $M_G\sim \sqrt\sigma_{adj}$; (2) The string tension follows the Casimir scaling, i.e. $\sigma_{adj}=\frac{C_2(adj)}{C_2(f)}\sigma$. In a constituent gluon picture, our results suggest a low-lying glueball spectrum made of two transverse constituent gluons bound by an adjoint string, completed by three transverse constituent gluons bound by a Y-junction of adjoint strings rather than a $\Delta-$shaped junction of fundamental strings. 
\end{abstract}

\maketitle

\section{Introduction}
\label{introduction}

After decades of research, glueballs remain challenging particles to understand, both from experimental and theoretical point of views. We refer the reader to \cite{Llanes-Estrada:2021evz} for a recent review. Regarding experimental detection of glueballs, significant progress has been made recently, with the discovery of the $X(2370)$ as a pseudoscalar glueball candidate  \citep{BESIII:2023wfi}, or with $pp$ and $p\bar p$ elastic scattering data compatible with the odderon \citep{D0:2020tig}, i.e. with $C-$odd gluonic states that could contain the $(odd-J\geq 3)^{--}$ glueballs \citep{PhysRevD.109.034007}. From the theoretical side, a milestone in glueball study has been the computation of SU(3) pure Yang-Mills spectrum in lattice QCD \citep{Morningstar:1999rf,Chen:2005mg, Athenodorou:2020ani}. Lattice calculations have been extended to $(3+1)-$dimensional Yang-Mills theories for SU($N$) \citep{Athenodorou:2021qvs} and Sp($2N$), and also to $(2+1)-$dimensional Yang-Mills theories for SO($2N$) \citep{Bursa:2012ab} although we here focus on $(3+1)-$dimensional cases \footnote{We only quote the latest lattice studies and refer the interested reader to the references of the latter.}.

The understanding of Yang-Mills spectrum with effective models (constituent models, AdS/QCD, QCD sum rules,\dots) has motivated a great number of studies, see e.g the review of \cite{Mathieu:2008me}. Although the SU(3) lattice spectrum has been widely studied, the generalization of effective approaches to other gauge algebras has been less considered with effective models. In previous works \citep{Buisseret:2011bg,Buisseret:2013ch} we have proposed a constituent-gluon approach of glueballs that can accommodate a change of gauge algebra by resorting to Casimir scaling for the main parameter of the theory, i.e. the string tension. We summarize the proposal in Sec. \ref{casimir} and apply it to the most recent lattice data in the SU($N$) and Sp($2N$) cases in Sec. \ref{fit}. Finally we compute SU($\infty$) and Sp($\infty$) glueball masses within a constituent approach in Sec. \ref{spectrum}

\section{Casimir scaling in glueballs}\label{casimir}

We assume that the Hamiltonian ${\cal H}$ generating glueball states depends on two main types of interactions: long-range nonperturbative interactions and short-range, one-gluon-exchange-type ones. The nonperturbative interactions depend on an energy scale $\mu$ and the short-range interactions depend on the product $C_2(adj)\alpha_S$, with $C_2(adj)$ the quadratic Casimir operator of the gauge algebra in the adjoint representation --  more generally $C_2(R) =  T_R\cdot T_R$ with $T_R$ the generators of the gauge algebra in the representation $R$. The coupling constant $\alpha_S=\frac{g^2}{4\pi}$, and renormalization arguments lead to $	g^2=\frac{\lambda}{C_2(adj)}$ with $\lambda$ the 't Hooft coupling \citep{tHooft:1973alw,Witten:1979kh} that we further assume to be gauge-algebra independent. Therefore, $C_2(adj)\alpha_S= \frac{\lambda}{4\pi}=\bar\alpha$, that we assume to be a gauge-independent parameter, and ${\cal H}={\cal H}(\mu,\lambda)$. Since $\mu$ is the only energy scale, any glueball mass $M_G$ should behave as $M_G=\mu\, f_G(\lambda)$, with $f_G(\lambda)$ a gauge-independent constant. The nonperturbative energy scale $\mu$ may be identified to the square root of the string tension in some representation $R$, that is $\sqrt{\sigma_R}$ the  tension of the string generated by a colour source in $R$. One has $M_G(R)=\sqrt{\sigma_R}\bar M_G$, where $\bar M_G$ is assumed to be gauge-algebra independent. Any choice of $R$ will result in a different prediction for the glueball mass behaviour when changing gauge-algebra. 

According to strong coupling expansion of lattice QCD, the string tension $\sigma_R$ is given by $\sigma_R =C_2(R)\, g^2 \Omega$ \citep{DelDebbio:2001sj}, where $\Omega$ is gauge-algebra independent. Therefore, one can write 
\begin{equation}
	\sigma_R=\frac{C_2(R)}{C_2(f)}\sigma,
\end{equation}
with $\sigma=\sigma_f$ the string tension. This is the Casimir scaling hypothesis (see \citep{Bali:2000un,Semay:2004br} for a numerical validation of this hypothesis on static sources in the SU(3) case). The Casimir scaling is also verified at next-to-next-to leading order \citep{Berw2016}. Finally, we are led to 
\begin{equation}\label{Mmodel}
	\frac{M_G(R)}{\sqrt\sigma}=\sqrt{\frac{C_2(R)}{C_2(f)}}\ \bar M_G\equiv \eta(R)\ \bar M_G,
\end{equation}
$\bar M_G$ being independent of the gauge algebra.

We will consider two scenarios. First, the case where $R$ is the fundamental representation leads to a constant ratio  $\frac{M_G(f)}{\sqrt\sigma}=\bar M_G$. Second, the case where $R$ is the adjoint representation leads to a non-constant ratio $	\frac{M_G(adj)}{\sqrt\sigma}= \eta(adj)\ \bar M_G.$ The difference between both cases relies on which type of string sets the global mass scale, the fundamental or adjoint string. In the SU($N$) and Sp($2N$) cases, where the $N\to\infty$ limit is worth being considered, our two proposals read
\begin{eqnarray}
	\frac{M_{G}(adj)}{\sqrt\sigma}&=&\frac{\eta(adj)}{\sqrt 2}\frac{M_\infty}{\sqrt\sigma},\label{model1}\\
	\frac{M_{G}(f)}{\sqrt\sigma}&=&\frac{M_\infty}{\sqrt\sigma},\label{model3},
\end{eqnarray}
where $M_\infty$ are the $N\to\infty$ masses. Let us recall that
\begin{eqnarray}\label{etadef}
	\eta(adj)&=&\sqrt{\frac{2N^2}{N^2-1}}\quad {\rm for}\quad {\rm SU}(N)\nonumber \\
	&=&\sqrt{\frac{4(N+1)}{2N+1}}\quad {\rm for}\quad {\rm Sp}(2N),
\end{eqnarray}
that $\eta(f)=1$ by definition and that $\eta(adj)\to \sqrt 2$ as $N\to\infty$. We also note that, $\eta(adj)=\frac{3}{2}$ for SU(3) and $\frac{4}{\sqrt 7}$ for Sp(6), that is a difference of less than 1$\%$.

\section{Comparison to lattice QCD}\label{fit}

We are now in position to check whether latest lattice data are compatible with one scaling or another. SU($N$) glueball masses, $M_G$, were computed in \cite{Athenodorou:2021qvs} for $N=$2, 3, 4, 5, 6, 8, 10, 12, and the $N\to\infty$ limit was obtained by a fit of the form $M_G=c_0+\frac{c_1}{N^2}$. Sp($2N$) glueball masses were computed by \cite{universe9050236} for $N=$1, 2, 3, 4, and the $N\to\infty$ limit was obtained by a fit of the form $M_G=c_0+\frac{c_1}{2N}$. We rather fit the finite-$N$ data by models (\ref{model1}) and (\ref{model3}) using least-square method. The result of the fit are the remaining parameter $M_\infty$ and the $\chi^2$ per degree of freedom, that we use as a assessment of the quality of the fit -- the closer to 1, the better the fit. The string tension $\sigma$ plays no role in the fit since we use normalized masses. When two lattice states are related to the same glueball state, i.e. the $T_2^\pm$ and $E^\pm$ channels for Sp($2N$) in \cite{universe9050236}, the fits are performed using both data simultaneously.

Results are presented in Table \ref{tab1} for SU($N$) and in Table \ref{tab1b} for Sp($2N$). We restrict our analysis to the fundamental $J^{PC}=(0,1,2,3)^{\pm\pm}$ channels since the errors on $4^{\pm\pm}$ (and higher-lying) states are so large that they prevent a clear interpretation. The SU($\infty$) and Sp($\infty$) masses are close to each other. Universality arguments state that they have actually to be equal \citep{Lovelace:1982hz}, although this constraint was not included in our regressions. 

\begin{table}[h]
	\caption{Fit of the SU($N$)lattice data of  \cite{Athenodorou:2021qvs} according to models  (\ref{model1}) and (\ref{model3}). For each fit, $M_\infty$ and $\chi^2$ per degree of freedom are displayed and the values between parenthesis are the standard deviations of $M_\infty$ according to least-square method. The best fit ($\chi^2$ closer to unity) is marked with a *.}
	\label{tab1}
	\begin{tabular}{c|cc|cc}
		&\multicolumn{2}{c|}{$M_{G}(adj)$}
		&  \multicolumn{2}{c}{$M_{G}(f)$}\\
		\hline
		$J^{PC}$ & $M_\infty$&  $\chi^2$ per d.o.f.   & $M_\infty$ &  $\chi^2$ per d.o.f. \\
		\hline
		$0^{++}$  & 3.144(28)   &10.7* & 3.258(84) &95.2  \\
		$0^{-+}$  & 4.875(64) &  11.0* &5.050(152) & 54.1\\ 
		$1^{-+}$&   8.189(84)   &  3.48*&8.504(89)  & 3.69\\
		$1^{+-}$  & 5.753(10) & 0.312*  & 5.872(43) & 4.98 \\
		$1^{--}$ & 7.393(95)  & 4.74*  &7.543(146) &11.7\\
		$2^{++}$  & 4.610(7) & 1.01* & 4.782(87) & 141 \\
		$2^{-+}$  & 6.020(16)  & 0.762* & 6.245(116)   & 39.7\\
		$2^{+-}$&   8.497(48) &  1.41*& 8.673(69) &2.03 \\
		$2^{--}$  & 7.754(60)   & 4.01  &7.916(53) & 2.77* \\
		$3^{++}$    &7.190(48)   &  1.92* & 7.508(147)  & 15.4 \\
		$3^{+-}$  & 6.917(28)  &  5.51* &7.060(45) & 49.1 \\
	\end{tabular}
\end{table}

\begin{table}[h]
	\caption{Same as Table \ref{tab1} for the Sp($2N$) lattice data of \cite{universe9050236}.}
	\label{tab1b}
	\begin{tabular}{c|cc|cc}
		&\multicolumn{2}{c|}{$M_{G}(adj)$}
		&  \multicolumn{2}{c}{$M_{G}(f)$}\\
		\hline
		$J^{P}$ & $M_\infty$ &  $\chi^2$ per d.o.f. & $M_\infty$ &   $\chi^2$ per d.o.f.\\
		\hline
		
		$0^{+}$  & 3.239(40)   & 1.05* & 3.539(115) &  7.07 \\
		$0^{-}$  & 5.143(103)  & 1.71* &5.618(218)   & 7.16 \\
		$1^{+}$ &  7.649(179)  & 0.830* & 8.368(204)  & 0.828 \\
		$1^{-}$   & 8.209(253)  &3.10 & 8.990(129) &    0.649* \\
		$2^{+}$   & 4.610(55)  &  0.456 & 5.040(116) & 1.16* \\ 
		$2^{-}$ & 6.000(130)   & 4.48 & 6.565(112)  &4.38*   \\
		$3^{+}$   & 6.527(312)  & 9.72 &7.145(311)  &  8.25*\\
		$3^{-}$   & 7.802(342)  & 1.54*  &8.523(464)  &  1.80\\
	\end{tabular}
\end{table}

\begin{figure}
	\centering
	\includegraphics[width=0.9\linewidth]{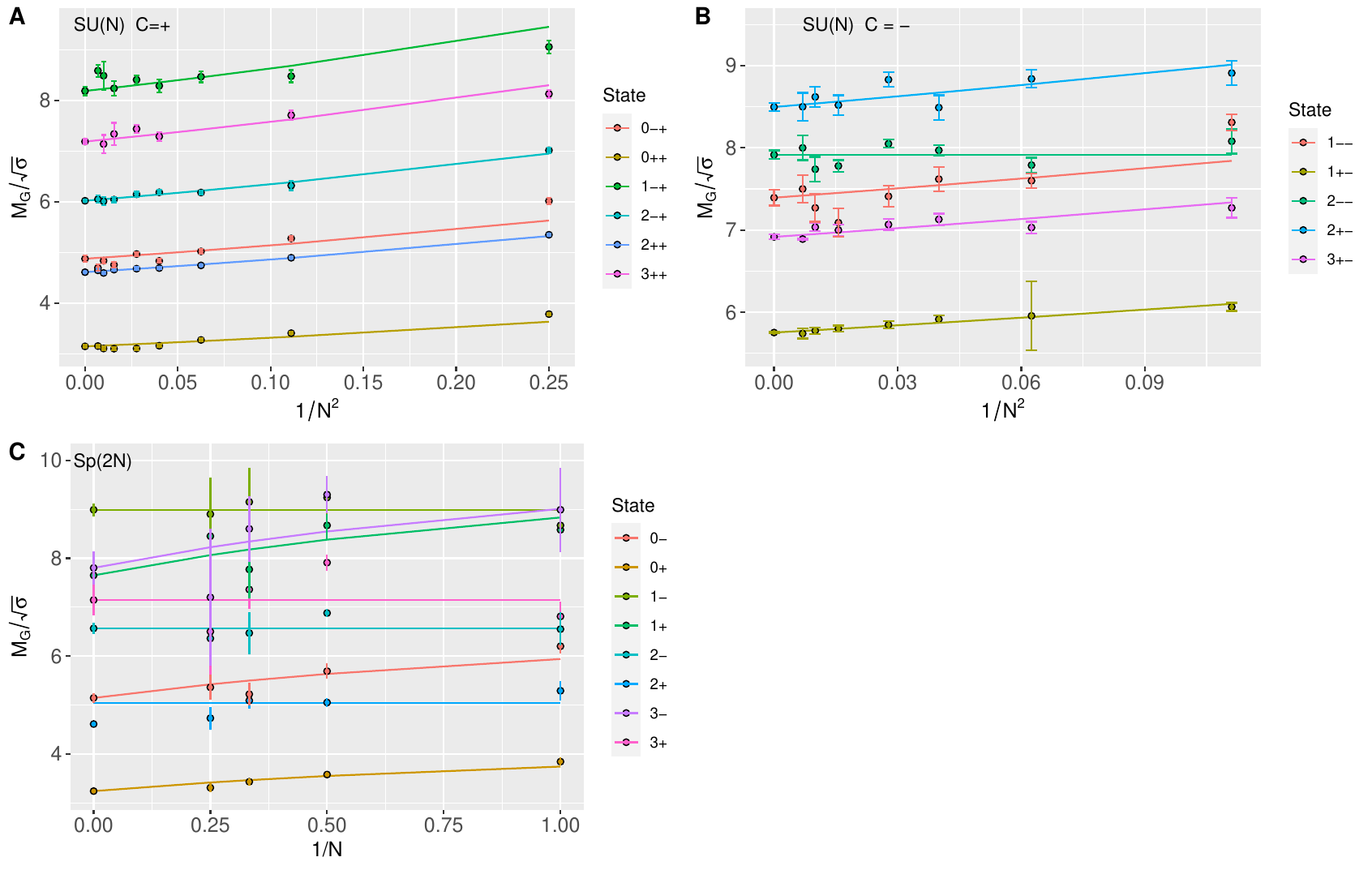}
	\caption{A: Evolution of $C=+$ SU($N$) glueball masses versus $1/N^2$. Lattice data of \cite{Athenodorou:2021qvs} (points) are compared to the best fits presented in Table \ref{tab1}. The $N\to\infty$ mass is the one obtained by our best fit. B: Same as panel A for $C=-$ SU($N$) glueballs. C: Same as panel A for Sp($2N$) glueballs but masses are plotted versus $1/N$; lattice data are taken from \cite{universe9050236}. For the sake of clarity, only the $T_2^\pm$ lattice states are plotted for the $2^\pm$ states.}
	\label{Fig1}
\end{figure}

The quality of the fits can be appraised in Fig. \ref{Fig1}. From Tables \ref{tab1} and \ref{tab1b}, it can be checked that 14 of 19 states are compatible with a scaling of type $M_G(adj)$, i.e. the adjoint string sets the scale of glueball masses. The situation is especially clear with the scalar and pseudoscalar states. In a constituent picture they can be interpreted as two-gluon states bound by an adjoint string. This conclusion is common to SU($N$) and Sp($2N$) gauge groups and exemplifies the conjecture of \cite{Hong:2017suj}, that has also been explicitly checked for SU($N$) and SO($2N$) in their paper.

More globally, the behaviour of all SU($N$) glueball masses are compatible with a scaling in $\sqrt{\sigma_{adj}}$, the $2^{--}$ state excepted. In a constituent gluon picture, $C=-$ states are at least three-gluon ones. Therefore, our results suggest that three-gluon glueballs interact via a Y-junction of adjoint strings rather than a $\Delta-$junction of fundamental strings. \cite{Cardoso} have previously carried a SU(3) lattice computation of the static energy between three colour octet sources located either on an equilateral triangle or on a isosceles rectangle triangle. They have found that, at large source separations, the static energy is proportional to the perimeter of the triangle, which is the case for both Y- or $\Delta$-junction (the former being seen as proportional to the half-perimeter). Moreover, they find a slope that is only compatible with a $\Delta$-junction assuming that $\sigma$ has the same value in both $\Delta$- and Y-junction. Our results are not in agreement with this conclusion but, as shown by \cite{Silvestre-Brac:2003ybk} in the baryon case, a rescaling of $\sigma$ by less than 10$\%$ may lead to  $\Delta$- and Y-junction static energies very close to each other.  Allowing $\sigma$ to be rescaled may allow both for $\Delta$- and Y-junction behaviours to be valid from the data of \cite{Cardoso}. In the present paper we focus on the colour scaling of the string tension for SU($N$) rather than on the static energy shape for SU(3), which leads us to the conclusion that adjoint strings, rather than fundamental ones, are favoured. Further lattice computations of the static energy of three adjoint sources for SU($N$) are needed to shed more light on the discrepancy between our results and those of \cite{Cardoso}. 

The higher-lying part of the Sp($2N$) spectrum (beyond $1^+$ state) is rather compatible with a scaling in $\sqrt{\sigma}$ and not $\sqrt{\sigma{adj}}$, although all fits are poorer in this part of the spectrum. $1^-$  glueball has to be interpreted as at least a three gluon state, so in this case the most favoured picture is that of three gluons bound by fundamental strings. However, we think that there is a need to improve the accuracy of lattice data for high-lying glueball states in Sp($2N$) to clarify the picture before validating a preferred model.

\section{Constituent model}\label{spectrum}

The two-body spinless Salpeter Hamiltonian
\begin{equation}
	H_{gg}=2\sqrt{\vec p^{\, 2}}+\sigma_{adj} r-C_2(adj)\frac{\alpha_S}{r}+\sqrt{\sigma_{adj}}\, C_{gg},
\end{equation}	
with $C_{gg}$ an arbitrary but gauge-algebra independent constant, can be rewritten as 
\begin{equation}\label{gg}
	\frac{H_{gg}}{\sqrt\sigma}=\eta(adj) \left[2\sqrt{\vec p^{\, 2}}+ r-\frac{\bar\alpha}{r}+C_{gg}\right]
\end{equation}	
after rescaling $\vec p\to \sqrt{\sigma_{adj}}\vec p$, $\vec x\to \vec x/\sqrt{\sigma_{adj}}$. It is a Hamiltonian that leads to the observed scaling for two-constituent gluon glueballs while accurately reproducing SU(3) lattice data provided constituent gluons are transverse helicity-1 particles \citep{Mathieu:2008bf}. We will use this Hamiltonian to model all $C=+$ glueballs but $1^{++}$ and $({\rm odd}-J)^{-+}$ ones. These quantum numbers are indeed forbidden for bound states of two helicity-1 particles, see also \cite{Boulanger:2008aj}.

The three-body spinless Salpeter Hamiltonian
\begin{eqnarray}
	H^Y_{ggg}&=&\sum^3_{i=1}\left[\sqrt{\vec p^{\, 2}_i}+\sigma_{adj} \vec x_i-\vec R\vert\right]-\sum^3_{i<j}\left[\frac{C_2(adj)}{2}\frac{\alpha_S}{\vert \vec x_i-\vec x_j\vert}\right]+\sqrt{\sigma_{adj}}\, C_{ggg}\\
	&\approx&\sum^3_{i=1}\sqrt{\vec p^{\, 2}_i}
	+\sum^3_{i<j}\left[\frac{\sigma_{adj}}{2}\vert \vec x_i-\vec x_j\vert  - \frac{C_2(adj)}{2}\frac{\alpha_S}{\vert \vec x_i-\vec x_j\vert}\right]+\sqrt{\sigma_{adj}}C_{ggg} ,
\end{eqnarray}
where we approximated the Y-junction by the half-perimeter (a correct approximation as shown in \cite{Silvestre-Brac:2003ybk}), also has the needed scaling:
\begin{eqnarray}\label{gggY}
	\frac{H^Y_{ggg}}{\sqrt\sigma}&=&\eta(adj)\left[\sum^3_{i=1}\sqrt{\vec p^{\, 2}_i}
	+\sum^3_{i<j}\left(\frac{1}{2}\vert \vec x_i-\vec x_j\vert  - \frac{1}{2}\frac{\bar\alpha}{\vert \vec x_i-\vec x_j\vert}\right)+C_{ggg}\right],
\end{eqnarray}
where $C_{ggg}$ is an arbitrary gauge-algebra independent constant. We use this Hamiltonian to compute the $C=-$ glueball spectrum as performed in \cite{Chevalier:2025xed}, to which we refer the reader for technical details about bound state wave functions of three-body helicity states. In the present work, we extend the spectrum to $C=+$ three-gluon states, i.e. totally antisymmetric states. It enables access to the lightest $1^{-+}$ and $3^{-+}$ glueballs. These masses are obtained using the same methodology as in \cite{Chevalier:2025xed}, but with $\sigma = -1$. For reproducibility, we note that the optimal non-linear variational parameters for $1^{-+}$ and $3^{-+}$ are $(a,b)=(0.55,2.3)$ and $(a,b)=(0.8,2.6)$ respectively, as defined by \cite{Chevalier:2025xed}.

\begin{figure}
	\centering
	\includegraphics[width=0.5\linewidth]{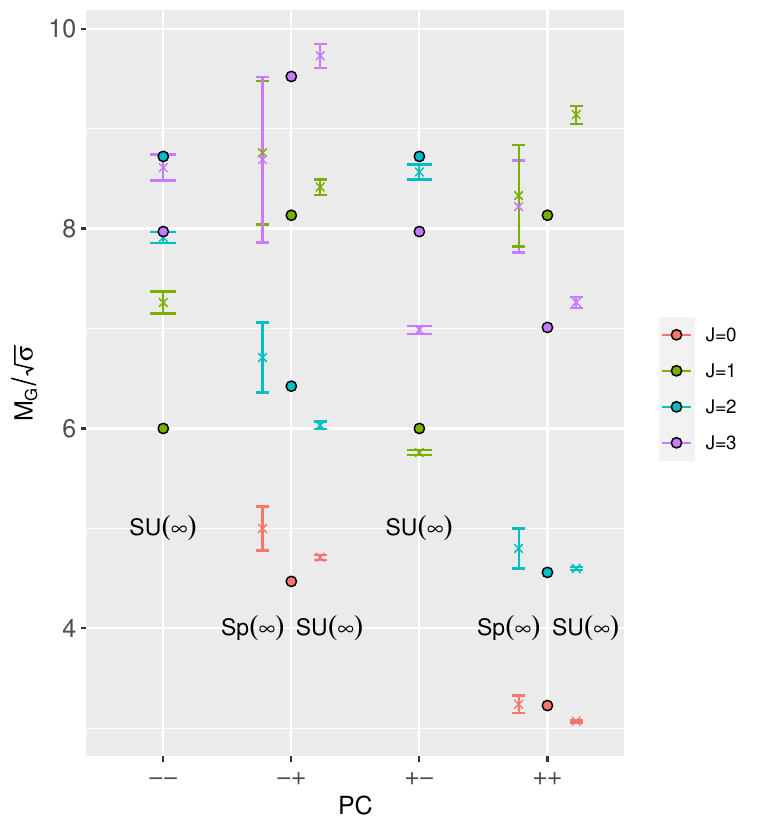}
	\caption{Constituent model (points) compared to lattice data (error bars) for SU($\infty$) \citep{Athenodorou:2021qvs} and Sp($\infty$) \citep{universe9050236}. Only the lightest states of each $J^{PC}$ are displayed.}
	\label{fig:spectrum_constituent}
\end{figure}

Our model predicts a common glueball spectrum for SU$(\infty)$ and Sp$(\infty)$, in agreement with \cite{Lovelace:1982hz}. The mass spectrum is displayed in Table \ref{tab2} and plotted in Fig. \ref{fig:spectrum_constituent}. The arbitrary constants have been fitted on the lattice data by resorting to the least-square method. Low-lying three-constituent gluons $0^{\pm-}$ states are absent because their wave function in momentum space have to be very asymmetrical, which prevents light states \citep{Boulanger:2008aj}. As in the SU(3) case discussed by \cite{Chevalier:2025xed}, the agreement is satisfactory in the two-gluon sector given the simplicity of the model. Globally, the mass hierarchy of the lightest states in all the $PC$ channels is well reproduced but the situation is less conclusive in the three-gluon sector. It is plausible that neglected interactions such as perturbative and non-perturbative spin-orbit, or other hyperfine terms, play a more important role in the three-gluon sector. The inclusion of such interactions is left for future studies. 

\begin{table}[h]
	\caption{SU($\infty$) and Sp($\infty$) lowest-lying glueball masses for each quantum number in our constituent model defined by Hamiltonians (\ref{gg}) and (\ref{gggY}) with $\eta(adj)=\sqrt 2$, $\bar\alpha=1.35$, $C_{gg}=-0.62$ and $C_{ggg}=-4.02$, compared to the best fits given in Tables \ref{tab1} and \ref{tab1b} and to the lattice data. Empty cells denote masses that were not available from lattice QCD. $N_g$ is the number of constituent gluons related to a given state.  }
	\label{tab2}
	\begin{tabular}{c|cc|cc|cc}
		
		& \multicolumn{2}{c|}{Constituent Model } &  \multicolumn{2}{c|}{Best fit} & \multicolumn{2}{c}{Lattice} \\
		\hline
		$J^{PC}$ & $N_g$ & Mass &SU($\infty$)&Sp($\infty$) & SU($\infty$)  & Sp($\infty$)  \\
		\hline
		$0^{++}$ & 2 &3.228 & 3.144(28) & 3.239(40) & 3.072(14) & 3.241(88) \\
		$0^{-+}$ &  2& 4.470 & 4.875(64) & 5.143(103)  &  4.711(26)& 5.00(22)  \\
		$1^{++}$ & 3  & 8.133& &7.449(179)  & $\geq$9.14(9)&8.33(51)  \\
		$1^{-+}$ & 3  & 8.133  & 8.189(84) & 8.990(129) & 8.415(76) & 8.76(72) \\
		$1^{+-}$ & 3 &  6.000  & 5.753(10) & & 5.760(25) &  \\
		$1^{--}$ & 3  &  6.000  & 7.393(95)& & 7.26(11)&  \\
		$2^{++}$ &  2  & 4.560  & 4.610(7)& 4.610(55)  & 4.599(14) & 4.80(20) -- $T_2^+$ \\
		&    &  & &   &  & 4.79(19) -- $E^+$ \\
		$2^{-+}$ & 2 & 6.423  &6.020(16)  & 6.000(130)& 6.031(38) & 6.71(35) -- $T_2^-$\\
		&  & &  & &  & 6.44(33) -- $E^-$\\
		$2^{+-}$ & 3 & 8.722 & 8.497(48)  &&8.566(76) &  \\
		$2^{--}$ & 3 & 8.722 &  7.916(53) &&7.910(56) &  \\
		$3^{++}$ 	& 2 & 7.011 & 7.190(48) & 7.145(311) &  7.263(56)&8.22(46)  \\
		$3^{-+}$ 	& 3 & 9.522 &  & 7.802(342) & $\geq$9.73(12) & 8.69(83) \\
		$3^{+-}$ 	& 3 &  7.970 & 6.917(28) &  & 6.988(41) &  \\
		$3^{--}$ 	& 3  & 7.970  &  &  &  $\geq$ 8.61(13)&  \\
		
	\end{tabular}
\end{table}

\section{Concluding comments}

In summary, we have shown that the ratio $\frac{M_G}{M_\infty}$ scales as $\sqrt{\frac{C_2(adj)}{C_2(f)}}$ in SU($N$) Yang-Mills theory and in the low-lying part of the Sp($2N$) Yang-Mills theory, $M_\infty$ being the $N\to\infty$ limit of the glueball masses. This result is in agreement with the previous findings of \cite{Hong:2017suj} and also with a similar scaling law based on the Casimir scaling in the mesonic sector \citep{Buisseret:2019ugy}. Such scaling laws may help to put constraints on effective models parameters, as we have shown in the case of constituent gluons approaches, where the observed scaling law favours adjoint strings as the best candidate to model interactions between constituent gluons. A simple  two-body spinless Salpeter Hamiltonian with funnel potential is able to model the SU($\infty$) and Sp($\infty$) $C=+$ low-lying spectrum provided constituent gluons are seen as transverse particles. A three-body $Y-$junction generalization of this Hamiltonian leads to a $C=-$ low-lying spectrum with the expected global features but showing extra states with respect to the lattice QCD spectrum, as already pointed out in \cite{Chevalier:2025xed}.

\bibliographystyle{apsrev}
\bibliography{Glueb_su_sp}



%% For example:

%% \bibitem[Aladro et al.(2015)]{Aladro15} Aladro, R., Martín, S., Riquelme, D., et al. 2015, \aas, 579, A101

%%\end{thebibliography}

\end{document}